\documentstyle[aps,preprint,epsfig,floats]{revtex}

\tighten

\newcommand{\half}{\frac{1}{2}}

\newcommand{\be}{\begin{equation}}
\newcommand{\ee}{\end{equation}}
\newcommand{\bea}{\begin{eqnarray}}
\newcommand{\eea}{\end{eqnarray}}

\newcommand{\BS}{Bethe--Salpeter }

\newcommand{\w}{\omega}

\newcommand{\s}{\!\cdot\!}

\newcommand{\ov}[1]{\overline{#1}}
\newcommand{\dj}{\,\,\,\raisebox{-0.4ex}{\large $\bar{}$}\!\!d\,}
\def\slr#1{\setbox0=\hbox{$#1$}           
   \dimen0=\wd0                                 
   \setbox1=\hbox{/} \dimen1=\wd1               
   \ifdim\dimen0>\dimen1                        
      \rlap{\hbox to \dimen0{\hfil/\hfil}}      
      #1                                        
   \else                                        
      \rlap{\hbox to \dimen1{\hfil$#1$\hfil}}   
      /                                         
   \fi}

\begin{document}
\date{\today}
\title{Mesons in a Poincare Covariant Bethe--Salpeter Approach}

\author{R.~Alkofer, P.~Watson, 
and H.~Weigel\thanks{Heisenberg--Fellow}}

\address{Institute for Theoretical Physics, T\"ubingen University\\
D-72076 T\"ubingen, Germany}

\maketitle

\begin{center}
hep-ph/0202053 \hskip2cm UNITU-HEP-4/2002
\end{center}

\begin{abstract}  
We develop a covariant approach to describe the low--lying scalar, 
pseudoscalar, vector and axialvector mesons as quark--antiquark bound states. 
This approach is based  on an effective interaction modeling of the
non--perturbative structure  of the gluon propagator 
that enters the quark Schwinger--Dyson and meson Bethe--Salpeter
equations. We consistently treat these integral equations by precisely
implementing the quark propagator functions that solve the Schwinger--Dyson
equations into the Bethe--Salpeter equations in the relevant kinematical region.
We extract the meson  masses and compute the pion and kaon decay constants.
We obtain a quantitatively correct description for pions, kaons and vector
mesons while the calculated spectra of scalar and axialvector mesons 
suggest that their structure is more complex than being 
quark--antiquark bound states.
\\
\end{abstract}
~\\
{\it PACS: 14.40.-n,11.10.St,12.38.Lg} \\

\newpage 

\section{Introduction}

Recently, the scalar mesons have attracted a lot of interest as the 
reanalysis of the pseudoscalar meson scattering data indicated the 
existence of a flavor SU(3) nonet in this channel~\cite{Syr1}. 
It is therefore desirable to gain deeper understanding of the constituent 
structure of the scalar mesons together with a comprehensive description
of the meson states in the other spin--parity channels. The ultimate 
goal would be to understand all low--lying meson states and resonances 
as non--perturbative bound states in Quantum Chromo Dynamics~(QCD). 

A relativistic framework for analyzing mesons as composite objects is  provided
by the Bethe--Salpeter equations that extract poles in the  quark--antiquark
scattering kernel\cite{alkofer00}. The attraction  needed to bind quarks and
antiquarks emerges from dressed multiple gluon  exchange. Thus the essential
ingredients to these equations are the quark and gluon propagators as well as
the quark--gluon vertex. In addition, these $n$-point Green's functions are
related by their  Schwinger--Dyson equations which are part of an infinite
tower of non--linear integral equations. There has been some progress in the
understanding of the infrared behavior of the gluon propagator from recent
Yang--Mills lattice calculations \cite{Mandula:1999nj} as well as from 
studies of the coupled system of gluon and ghost Schwinger--Dyson equations 
\cite{alkofer00,vonSmekal:1997is}. Nevertheless, for phenomenological
applications the frequently adopted strategy is to model the gluon 
propagator as well as the quark--gluon vertex and consistently derive 
the quark propagator from its Schwinger--Dyson equation because it
facilitates the continuation to complex Euclidean momenta.  

These types of calculations have a long history, for reviews see
Refs.~\cite{alkofer00,roberts00}. Early versions adopted pointlike
gluon propagators in coordinate space that eventually lead to
Nambu--Jona--Lasinio (NJL) type models~\cite{nambu61,ebert86}, 
pointlike propagators in momentum space were also considered~\cite{Ja93}.
These models are particularly simple because either solving the
Schwinger--Dyson equation yields a free quark propagator or the
Bethe--Salpeter integral equations reduce to algebraic equations. 
The main target particularly of the NJL--model studies have
been the pseudoscalar mesons. It turned out that they can be
adequately described once the important feature of dynamical
chiral symmetry breaking is incorporated, {\it i.e.} the interaction 
is strong enough so that the resulting quark propagator signals a
non--vanishing quark condensate reflected by a non--zero constituent 
quark mass. Then the pseudoscalar mesons can be understood as the 
{\it would--be} Goldstone bosons of chiral symmetry breaking. However, 
as the Bethe--Salpeter equations involve the dressed (constituent) quark 
propagators, binding can only be achieved kinematically and meson 
states with masses larger than twice the constituent quark mass 
cannot be described consistently. For that reason, model gluon 
propagators have been developed that yield quark propagators without 
poles for real momenta as an attempt to include the confinement 
phenomena~\cite{maris98,maris99}. Again these studies focused on 
pseudoscalar mesons~\cite{Ro96,Ma97} while a comprehensive investigation 
for the scalar, pseudoscalar, vector and axialvector mesons has not 
been carried out so far. Other studies~\cite{maris98,maris99}
made contact with perturbative QCD by considering a model gluon 
propagator that matches the pertinent anomalous dimension. 
This contribution has negligible effect on the meson 
properties, but its inclusion makes cumbersome the extraction of the
solutions to the Schwinger--Dyson equations for the large time--like 
momenta that enter the Bethe--Salpeter equations with meson states other 
than the pseudoscalars. We consider this attempt interesting, but as
we do not want to focus on the ultraviolet properties of mesons here, 
we regard it an unnecessary technical complication. Therefore we
focus rather on establishing a model as simple as possible that we 
consider a pertinent starting point to study the structure and properties 
of low--lying mesons in a relativistic framework. Our model interaction 
is parameterized in form of a non--trivial gluon propagator that 
contains sufficient strength to cause dynamical chiral symmetry breaking. 
For technical reasons it
turns out that a Gaussian shape function for the propagator in
momentum space is most suitable. Essentially we consider this model 
propagator as an effective interaction that relativistically describes 
the binding of quarks and antiquarks to mesons. Furthermore, we take 
the quark--gluon vertex function to be the tree level one since this 
procedure provides a framework that is consistent with chiral symmetry 
when the ladder approximation for the Bethe--Salpeter equation is 
employed \cite{alkofer00,roberts00}. For approaches going beyond
ladder approximation see e.g.~Ref.~\cite{Be96}. 

This paper is organized as follows. In Section II we will introduce the
effective interaction and solve the Schwinger--Dyson equation for the
quarks. We will put particular emphasis on the analytic continuation
of the resulting quark propagator to time--like momenta that enter 
the Bethe--Salpeter equations. We will discuss the structure of the
Bethe--Salpeter equations and
present the solutions in Section III. In Section IV we will present the 
numerical results in the sector of the three light quarks (up, down 
and strange) and compare them to the empirical data. In Section V
we will conclude and suggest a possible extension of the current
approach in particular with regard to the possibility that the 
scalar meson might have to be considered as two--quark -- two--antiquark
bound states~\cite{Syr1,Ja77}. We devote an appendix to discuss the
numerical stability of our results.

\section{The Quark Schwinger-Dyson Equation}

The purpose of this section is twofold. First we will explain our 
model gluon propagator whose dressing parameterizes the effective 
interaction and then solve the corresponding Schwinger--Dyson equation 
for the quark propagator.

As already discussed in the introduction we take a Gaussian form for 
dressing the model gluon propagator. This follows the work of 
Ref.~\cite{maris99} with the exception that we omit the logarithmic 
tail that matches perturbative QCD because we are not interested in 
the ultraviolet properties of mesons. We therefore write,
\be
g^{2}G_{\mu\nu}^{ab}(q)=4\pi^2 D \delta^{ab}t_{\mu\nu}(q)
\frac{q^2}{\w^2}\, \exp{\left(-\frac{q^{2}}{\w^{2}}\right)}
\label{eq:gluon}
\ee
where $\mu,\nu$ are Lorentz indices, $t_{\mu\nu}(q)$ is the transverse 
momentum projector and $a,b$ label color. While the prefactors in 
eq.~(\ref{eq:gluon}) are chosen to make subsequent equations more 
concise,~$D$ and~$\w$ are dimensionful parameters that we will determine 
from fitting empirical data. The coefficient~$D$ sets the strength of the 
interaction and~$\omega$ is the value at which the scalar function in 
the parameterization is maximal. Hence~$\omega$ sets the interaction scale.  
The dressed gluon propagator~(\ref{eq:gluon}) is supposed to represent 
a sensible hadron model and hence one can envisage that~$\omega$ will have 
a value of several hundred~${\rm MeV}$. 

We interpret the effective interaction~(\ref{eq:gluon}) as the propagator 
(in Landau gauge) of a gluon that gets absorbed and emitted by the 
quarks that eventually get bound to form mesons. To completely define 
the interaction, we need to parameterize the quark--gluon coupling. 
To establish chiral symmetry we apply the rainbow--ladder approximation 
to the system of Schwinger--Dyson and Bethe--Salpeter equations. 
This implies that the quark--gluon coupling is given by the tree level
interaction vertex, $ig\gamma_\mu \frac{\lambda^a}{2}$, where 
$\lambda^a$ is a Gell--Mann matrix acting in color space. Note, that 
we have already included the coupling constant $g$ in the definition of 
the effective interaction~(\ref{eq:gluon}).

Then the Schwinger--Dyson equation for the (inverse) quark propagator 
becomes ($\dj^{4}k=d^{4}k/(2\pi)^{4}$)
\be
S^{-1}(p)=i\slr{p}+m_0+\int\dj^{4}k\,
\gamma_{\mu} S(k)\gamma_{\nu}\, g^2\, 
\frac{\lambda^a}{2}\frac{\lambda^b}{2}\,G_{\mu\nu}^{ab}(k-p)
\label{eq:qdse}
\ee
where $m_0$ is the current mass of the considered quark. This
contribution represents the only explicit distinction between quarks 
of different flavors. Of course, its effects will implicitly propagate 
through the whole calculation. However, for notational simplicity
we will continue to suppress flavor labels. A suitable parameterization 
of the quark propagator is inspired by the form of a free fermion propagator
\be
S(p)=\left[\frac{1}{i\slr{p}A(p^2)+B(p^2)}\right]\,.
\ee
In solving the Schwinger--Dyson equation~(\ref{eq:qdse}) we have to 
find the scalar functions $A(p^2)$ and $B(p^2)$. It is also very 
instructive to define a mass function via $M(p^2)=B(p^2)/A(p^2)$. In 
particular $M(p^2=0)$ plays the role of a constituent quark mass and a
large value thereof signals dynamical chiral symmetry breaking.
 
As usual we work in Euclidean space with Hermitian Dirac 
matrices\footnote{This can be related to the standard Minkowski space 
Dirac matrices via $\gamma_4=\gamma^M_0,\gamma_j=-i \gamma_j^M$.}
that obey $\{\gamma_{\mu},\gamma_{\nu}\}=2\delta_{\mu\nu}$ and
$\gamma_5=-\gamma_1\gamma_2\gamma_3\gamma_4$.  
Inserting the effective interaction~(\ref{eq:gluon}) and performing
the standard trace algebra, we then deduce the following coupled 
equations for the propagator functions 
\bea
A(x) &=& 1+\frac{D}{\w^2}\int_{0}^{\infty}
\frac{dy\,y\,A(y)}{(yA^2(y)+B^2(y))}
\nonumber\\&& \hspace{0cm}\times
\frac{2}{\pi}\int_{-1}^{1}dz\sqrt{1-z^2}
\left[-\frac{2}{3}y+\left(1+\frac{y}{x}\right)
\sqrt{xy}\,z -\frac{4}{3}yz^2\right]
\exp{\left\{-\frac{x+y-2\sqrt{xy}\,z}{\w^{2}}\right\}},
\label{eq:qdsea}
\eea
\bea
B(x)&=& m_0+\frac{D}{\w^2}\int_{0}^{\infty}
\frac{dy\,y\,B(y)}{(yA^2(y)+B^2(y))}
\nonumber\\&& \hspace{2cm}\times\frac{2}{\pi}
\int_{-1}^{1}dz\sqrt{1-z^2}\left[x+y-2\sqrt{xy}\,z\right]
\exp{\left\{-\frac{x+y-2\sqrt{xy}\,z}{\w^{2}}\right\}},
\label{eq:qdseb}
\eea
where the four dimensional integral measure has been expanded such 
that $x=p^2, y=k^2$ and $z=p\s k/\sqrt{p^2k^2}$.

In a first step we solve eqs.~(\ref{eq:qdsea}) and~(\ref{eq:qdseb}) 
for spacelike momenta, {\it i.e.} for real positive~$x$. Then we observe 
that the integrals on the $RHS$ of these equations only involve the 
propagator functions at real arguments $y$ and we can use them to 
numerically compute the propagator functions for {\it arbitrary 
complex} $x$.  At first sight, it appears that $A(x)$ and
$B(x)$ could not be consistently continued because the cut
along the negative $x$--axis (associated with $\sqrt{x}$) 
would yield different results when continuing in the 
upper or the lower half--plane and it would be impossible to 
resolve the ambiguity in $\sqrt{x}\to\pm i\sqrt{\xi}$ when 
continuing $x\to -\xi$. Fortunately, this is not an obstacle. 
By expanding the exponential functions in eqs.~(\ref{eq:qdsea}) 
and~(\ref{eq:qdseb}) it becomes obvious that all terms that 
contain $\sqrt{x}$ are odd in $z$ and thus vanish when
integrating over this angular variable. Thus we are free to 
choose either of the two signs above. For definiteness we
work in the upper half--plane with $\sqrt{x}\to i\sqrt{\xi}$
along the negative half--line.

The observation that it is sufficient to know the propagator functions 
along the real positive axis to compute them for all complex arguments
of interest 
is a major point of this present study. It is an especially important 
issue to the study of the Bethe--Salpeter equation, since the quark 
propagator must be sampled in the complex plane of Euclidean momenta. 
In other words, the quark functions are calculated at precisely 
the momenta for which they are used, with no fitting functions,
interpolation or extrapolation. However, in order to do so numerically,
the quark propagator functions must be known very accurately along 
the real positive axis. This is achieved by noting that in 
eqs.~ (\ref{eq:qdsea}) and~(\ref{eq:qdseb}) the angular ($z$) integrals 
can be done analytically for real and positive~$x$. This property of the
angular integrals is a 
particular feature of the Gaussian dressing function~(\ref{eq:gluon}).
The resulting equations read
\bea
A(x) &=& 1+D\int_{0}^{\infty}\!\frac{dy\,y\,A(y)}{(yA^2(y)+B^2(y))}
\exp{\left\{-\frac{x+y}{\w^{2}}\right\}}\nonumber\\&&
\hspace{2cm}\times
\left\{\left(1+\frac{y}{x}+2\frac{\w^2}{x}\right)
I_2\left(\frac{2\sqrt{xy}}{\w^2}\right)-2\frac{\sqrt{y}}{\sqrt{x}}I_1
\left(\frac{2\sqrt{xy}}{\w^2}\right)\right\},
\nonumber\\
B(x)&=& m_0+D\int_{0}^{\infty}\!\frac{dy\,y\,B(y)}{(yA^2(y)+B^2(y))}
\exp{\left\{-\frac{x+y}{\w^{2}}\right\}}
\nonumber \\ && \hspace{2cm}\times
\left\{\left(\frac{\sqrt{y}}{\sqrt{x}}
+\frac{\sqrt{x}}{\sqrt{y}}\right)I_1
\left(\frac{2\sqrt{xy}}{\w^2}\right)-2I_2
\left(\frac{2\sqrt{xy}}{\w^2}\right)\right\}\, ,
\label{eq:qdses}
\eea
where $I_n$ are modified Bessel functions. These equations are  
one--dimensional coupled non--linear integral equations which can be 
straightforwardly solved numerically with a high degree of precision. The 
solutions for real (Euclidean) momenta are 
subsequently substituted into eqs.~(\ref{eq:qdsea}) and~(\ref{eq:qdseb})
to yield the propagator functions for complex momenta.

\section{The Bethe-Salpeter Equation}

Having obtained the quark propagators in the complex plane from the 
Schwinger--Dyson equations we have collected all ingredients for the 
Bethe--Salpeter integral equations. They will ultimately yield the quark meson 
vertex functions, $\Gamma$, that describe mesons as bound quark--antiquark 
pairs. Strictly speaking, the \BS equation is an eigenvalue problem, 
valid only at the resonance pole $P^2=-M^2$, where $M$ is the mass of 
the resonance. It is derived from considering the four--point quark
Green's function that involves exchanges of resonance mesons.
Such an exchange is characterized by a pole in that four--point function,
and the homogeneous Bethe--Salpeter equation below~(\ref{eq:bse}) determines
the position of that pole. All other regular terms in the vicinity of 
this pole are neglected. Demanding furthermore that the residue of this 
pole is unity yields the normalization condition~(\ref{norm}) for the 
vertex functions.

\subsection{Bethe--Salpeter vertex functions}

The vertex functions resulting from the Bethe--Salpeter equation
are characterized by three momenta out
of which only two are linearly independent due to momentum conservation
at the vertex. If we denote the meson momentum $P$ and the momentum
of the incoming quark $p+\xi P$ then the momentum of the outgoing
quark (= incoming antiquark) is $p+(\xi-1)P$. This suggests to label
the vertex functions by $p$ and $P$: $\Gamma(p,P)$. We have also 
introduced the arbitrary momentum partition parameter~$\xi\in[0,1]$. 
Due to strict relativistic covariance the results for physical
observables do not depend on $\xi$. Unfortunately we will have to assume
approximations within the full numerical computation (see Section 4)
that violate covariance to some extent. We will study the 
$\xi$--dependence of our results and verify that within a wide range
$\xi$--independence is maintained, see Appendix A. This will represent an 
{\it a posteriori} validation for the relativistic covariance of our 
computations. 

We now turn to the main target of 
our studies, the Bethe--Salpeter integral equations for the vertex 
function~$\Gamma(p,P)$ in ladder 
approximation~\cite{alkofer00,roberts00,tandy}:
\be
\Gamma(p;P)=-\frac{4}{3}\int\dj^{4}k
\left[\gamma_{\nu}\,S(k+\xi P)\,\Gamma(k;P)\,S(k+(\xi-1)P)\,
\gamma_{\mu}\right]\,g^2 G_{\mu\nu}(k-p)\,.
\label{eq:bse}
\ee
Here we have factorized the color factors in the effective interaction, 
$G_{\mu\nu}^{ab}(q)=\delta^{ab}G_{\mu\nu}(q)$ and performed the 
corresponding trace. The flavor content of the meson is not made 
explicit in eq.~(\ref{eq:bse}) as we have suppressed the 
flavor labels in the quark propagators. It is understood 
that the two propagators in eq.~(\ref{eq:bse}) are taken 
such as to account for the flavor quantum numbers of the 
considered meson. 
In the model that we will consider, the up and 
down quarks will be assumed to have equal current masses ($m_0$ in 
eq.~(\ref{eq:qdseb})) and thus also identical 
propagator functions $A(x)$ and $B(x)$. For the light quarks, which 
should give rise to the familiar $SU(3)_f$ nonet, we are thus left with 
three representatives of each of the multiplets that are distinguished 
by their isospin number, $I=0,\half,1$ . We must also specify the
meson angular momentum and parity. This is reflected by the 
Dirac and Lorentz decomposition of the meson vertex functions. This 
decomposition is known in the literature and
here we follow Ref.~\cite{llew}. For the pseudoscalar 
channel~($J^{P}=0^{-}$) we take
\be
\Gamma^{(P)}(p;P)=\gamma_{5}\left[\Gamma^{(P)}_0(p;P)
-i\slr{P}\Gamma^{(P)}_1(p;P)-i\slr{p}\Gamma^{(P)}_2(p;P)
-\left[\slr{P},\slr{p}\right]\Gamma^{(P)}_3(p;P)\right]\, .
\label{eq:pseu}
\ee
The decomposition for a scalar ($J^{P}=0^{+}$) meson reads
\be
\Gamma^{(S)}(p;P)=\Gamma^{(S)}_0(p;P)-i\slr{P}\Gamma^{(S)}_1(p;P)
-i\slr{p}\Gamma^{(S)}_2(p;P)
-\left[\slr{P},\slr{p}\right]\Gamma^{(S)}_3(p;P).
\label{eq:scal}
\ee
The vector ($J^{P}=1^{-}$) channel involves eight scalar functions
\bea
\Gamma^{(V)}_{\mu}(p;P) &=& 
\left[\gamma_{\mu}-\frac{P_{\mu}\slr{P}}{P^2}\right]
\left[i\Gamma^{(V)}_0(p;P)+\slr{P}\Gamma^{(V)}_{1}(p;P)
-\slr{p}\Gamma^{(V)}_2(p;P)+i\left[\slr{P},\slr{p}\right]
\Gamma^{(V)}_3(p;P)\right]\nonumber\\
&&+\left[p_{\mu}-\frac{P_{\mu}p\s P}{P^2}\right]
\left[\Gamma^{(V)}_{2}(p;P)+2i\slr{P}\Gamma^{(V)}_3(p;P)\right]
\label{eq:vect}  \\ &&
+\left[p_{\mu}-\frac{P_{\mu}p\s P}{P^2}\right]
\left[\Gamma^{(V)}_4(p;P)+i\slr{P}\Gamma^{(V)}_5(p;P)
-i\slr{p}\Gamma^{(V)}_6(p;P)+\left[\slr{P},\slr{p}\right]
\Gamma^{(V)}_7(p;P)\right]\, .
\nonumber
\eea
In the axialvector channel we have two modes that are distinguished
by their charge conjugation properties~\cite{llew}. For $J^{PC}=1^{++}$ the
decomposition is
\bea
\Gamma^{(A)}_{\mu}(p;P) &=& 
\gamma_{5}\left[\gamma_{\mu}-\frac{P_{\mu}\slr{P}}{P^2}\right]
\left[i\Gamma^{(A)}_0(p;P)+\slr{P}\Gamma^{(A)}_{1}(p;P)
-\slr{p}\Gamma^{(A)}_2(p;P)+i\left[\slr{P},\slr{p}\right]
\Gamma^{(A)}_3(p;P)\right]
\nonumber\\ &&
+\gamma_{5}\left[p_{\mu}-\frac{P_{\mu}p\s P}{P^2}\right]
\left[\Gamma^{(A)}_{2}(p;P)+2i\slr{P}\Gamma^{(A)}_3(p;P) \right]\, ,
\label{eq:axip}
\eea
while the $J^{PC}=1^{+-}$ mode is decomposed as 
\be
\Gamma^{(\tilde{A})}_{\mu}(p;P) = 
\gamma_{5}\left[p_{\mu}-\frac{P_{\mu}p\s P}{P^2}\right]
\left[\Gamma^{(\tilde{A})}_1(p;P)+i\slr{P}\Gamma^{(\tilde{A})}_2(p;P)
-i\slr{p}\Gamma^{(\tilde{A})}_3(p;P)+\left[\slr{P},\slr{p}\right]
\Gamma^{(\tilde{A})}_4(p;P)\right].
\label{eq:axim}
\ee
In what follows we will omit the superscripts that label the 
spin and parity channels because these channels do not mix 
and there should hence be no confusion. 

The primary object of this paper is to extract the bound state masses
for the various flavor combinations and angular momentum channels.\footnote{  
The traceology that is involved to project the Bethe--Salpeter
equation (\ref{eq:bse}) onto the Dirac components $\Gamma_i$ is 
lengthy, but straightforwardly performed using the algebraic 
manipulation package FORM~\cite{form}.} The corresponding projection
results in sets of coupled equations for the $\Gamma_i$. After carrying out
two of the three angular integrals analytically we are left with 
functions of the squared momenta $p^2$ and $P^2$ as well as the
angle between $p$ and $P$: $z=p\s P/\sqrt{p^2P^2}$. The $z$--dependence
is analyzed by an expansion in Chebyshev polynomials $T_k$
\be
\Gamma_i(p;P)=\sum_{k=0}^{N_{ch}-1}(i)^k
\Gamma_i^k(p^2;P^2)T_k(z)\, .
\label{eq:cheb}
\ee
Since the $T_k$ form an orthonormal set, we can project the equations 
for the Dirac components onto $\Gamma_i^k(p^2;P^2)$. Finally, the 
$k^2$--integral in the Bethe--Salpeter equation~(\ref{eq:bse}) is 
implemented numerically as a matrix equation for the unknown 
$\Gamma_i^k(p_j^2;P^2)$, $p_j^2$ being the discrete values of the 
momentum squared. The kernel, $K$ of that matrix parametrically depends 
on the meson momentum $P^2$. We solve that matrix equation as an 
eigenvalue problem by tuning the meson momentum to $P^2=-M^2$, such
that ${\rm Det}(1-K)=0$. This yields the desired meson mass~$M$.

For practical reasons we need to truncate the expansion~(\ref{eq:cheb})
at a certain order, $N_{ch}$. As noted before this violates 
relativistic covariance. Fortunately the momentum partition
dependence provides a measure for the degree at which covariance
is violated. We can mitigate this degree by increasing $N_{ch}$.
Numerically we furthermore scan the momentum partition dependence of
the computed meson masses to {\it a posteriori} verify covariance for 
our solutions to the Bethe--Salpeter equations, see Appendix A.
Unless otherwise stated we use $N_{ch}=4$.

\subsection{Pseudoscalar leptonic decay constant}

The solution to the Bethe--Salpeter equation not only yields the
meson masses but also the meson quark vertex functions that
can be used to compute meson properties. Here we will focus
on the pseudoscalar decay constants $f_{\pi}$ and $f_K$. In order 
to calculate these, we first have to normalize the vertex functions 
$\Gamma(p;P)$.  The \BS equation is a homogeneous equation, and 
thus needs an additional normalization condition. As mentioned 
previously, that condition is obtained from demanding the pole in 
the four--quark Green's function to be unity. For equal momentum 
partitioning, ({\it i.e.} for $\xi=1/2$ only) it reads~\cite{tandy}
\bea
2P_{\mu}=3\int\dj^{4}k\,&&{\rm Tr}\left\{
\ov{\Gamma}(k,-P)\frac{\partial S(k+P/2)}{\partial P_{\mu}}
\Gamma(k,P)S(k-P/2)\right.\nonumber\\&&\left.
+\ov{\Gamma}(k,-P)S(k+P/2)\Gamma(k,P)
\frac{\partial S(k-P/2)}{\partial P_{\mu}}\right\}
\label{norm}
\eea
where the trace is over Dirac matrices.  The conjugate 
vertex function $\ov{\Gamma}$ is defined as
\be
\ov{\Gamma}(p,-P)= C\Gamma^T(-p,-P)C^{-1}\, ,
\ee
where use is made of the charge conjugation matrix, 
$C=-\gamma_2\gamma_4$. The quark propagator derivatives are 
calculated by differentiating the quark Schwinger--Dyson
equations~(\ref{eq:qdsea}) and~(\ref{eq:qdseb}) analytically and 
then numerically integrating the corresponding expressions.

The decay constants are finally obtained from the coupling of
the axial current to the quark loop~\cite{tandy}
\be
f=\frac{3}{M^2}\int\dj^{4}k\,{\rm Tr}
\left\{\Gamma(k,-P)S(k+P/2)\gamma_5\slr{P}S(k-P/2)\right\},
\ee
where again the trace is over the Dirac matrices and the vertex 
functions are normalized according to eq.~(\ref{norm}). Again we have
not made explicit the flavor quantum numbers.

\section{Numerical Results and Discussion}

We are now in the position to present our numerical results to
the system of Schwinger--Dyson and Bethe--Salpeter equations. 
Even though the meson results are the primary target of our
investigation it is fruitful to first consider the quark propagator
functions obtained from the Schwinger--Dyson equations, as they enter 
the kernel of the Bethe--Salpeter equations.

\subsection{Quark propagator functions}

\begin{figure}
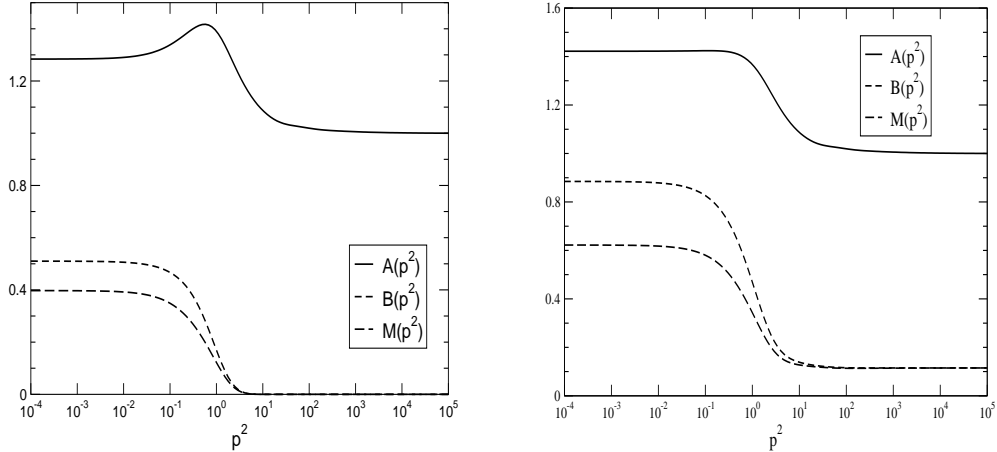

\centerline{
\epsfig{figure=quarks0.eps,width=6cm,height=6cm}\hspace{1cm}
\epsfig{figure=quarks1.eps,width=6cm,height=6cm}}
~\smallskip
\caption{Plot of the spacelike chiral quark functions (as a
function of the momentum squared).  The parameters are
$\w=0.5$GeV, $D=16.0$GeV$^{-2}$. Left panel: $m_0=0$, right panel: $m_0=0.115$
GeV.
All units are ${\rm GeV}$.}
\label{fig:quark1}
\end{figure}

In Fig.\ \ref{fig:quark1} we show the quark propagator functions 
$A(p^2)$, $B(p^2)$ and $M(p^2)=B(p^2)/A(p^2)$ along the positive, real 
spacelike axis,  $p^2>0$.  
This is the numerical 
solution to the coupled equations (\ref{eq:qdses}), which we emphasize 
is the basis for the quark solutions for complex momenta.  The solution 
clearly shows that dynamical chiral symmetry breaking is occurring: the 
mass function $M(p^2)$ attains a sizable non--zero value, even in the
case that the bare quark mass $m_0$ is zero. This phenomenon has been 
extensively studied (see for example Refs.~\cite{alkofer00,roberts00}), 
and is an example of genuinely non--perturbative behavior as dynamical 
mass generation cannot occur at any order in perturbation theory.  
Recall that the effective interaction (\ref{eq:gluon}) that enters the 
Schwinger--Dyson equations does not contain the perturbative UV behavior, 
rather it has an exponential damping at high momenta. This is manifested 
in the quark propagator functions as a sharp transition from the low 
momentum behavior to the bare values in the high momentum region. This 
transition occurs at about $1{\rm GeV}$.  

The quark mass function $M(p^2)$ is also a useful quantity to discuss 
the changes of the quark propagator with the input parameters. 
Table~\ref{tab:quark1} shows $M(p^2=0)$ for those parameter sets of 
$\omega$, $D$ and $m_0$ that we will later use to compute the meson masses. 
\begin{table}
\begin{center}
\begin{minipage}[t]{7.0cm}
\begin{tabular}{|c|c|c|c|}
$\w$ & $D$ & $m_0$ & $M(p^2=0)$ \\ \hline
$0.4$ & $45.0$& 0 & 0.520 \\
&        & $5\times10^{-3}$ & 0.531 \\
&        & 0.12 & 0.709 \\ \hline
$0.45$ & $25.0$& 0 & 0.450 \\
&        & $5\times10^{-3}$ & 0.462 \\
&        & 0.12 & 0.657 \\ \hline
$0.5$ & $16.0$& 0 & 0.397 \\
&        & $5\times10^{-3}$ & 0.413 \\
&        & 0.115 & 0.622
\end{tabular}
\end{minipage}
\end{center}
\caption{\label{tab:quark1}
The variation of the quark mass function evaluated at zero
momentum squared. All units are ${\rm GeV}$.}
\end{table}
As already mentioned
the quark mass function indicates the extent to which dynamical chiral 
symmetry breaking occurs and plays the role of the mass of the constituent 
quarks within hadrons. We see that the quark mass function decreases
as the effective scale $\omega$ increases and the strength $D$ decreases. 

\begin{figure}
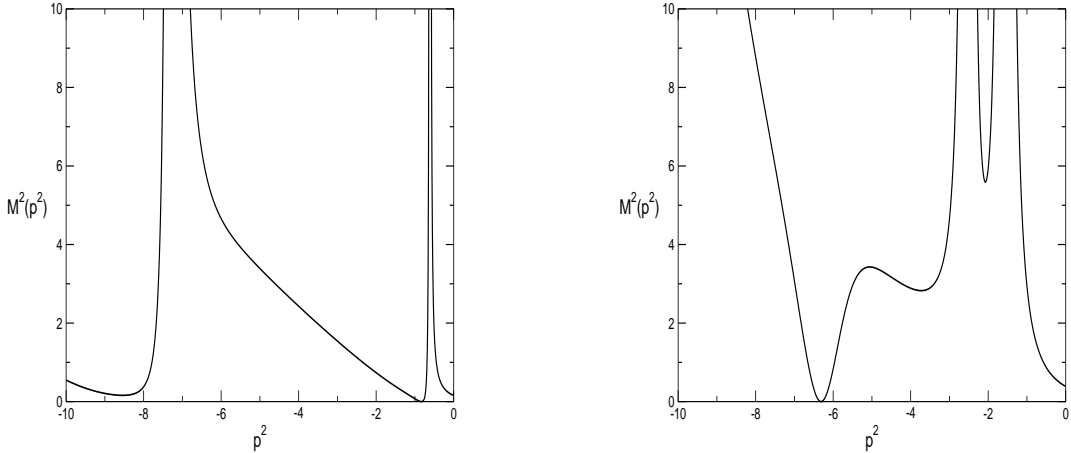

\centerline{
\epsfig{figure=quarkt0.eps,width=6cm,height=6cm}\hspace{2cm}
\epsfig{figure=quarkt1.eps,width=6cm,height=6cm}}
~\smallskip
\caption{Plot of the timelike quark mass function squared $M^2(p^2)$.
The parameters are $\w=0.5, D=16.0$. Left panel: $m_0=0$, 
right panel: $m_0=0.115$. All units are ${\rm GeV}$.}
\label{fig:quarkt}
\end{figure}

Since the main thrust of the paper is the use of the quark propagator 
functions evaluated at precisely the complex momenta squared for which 
they enter the \BS equation, it is important to study the behavior of 
the mass function at least along the timelike axis, {\it i.e.} for 
negative real momenta squared. In Fig.~\ref{fig:quarkt} we present 
typical results for the mass function squared $M^2(p^2)$ in that regime.
Although the effective interaction (\ref{eq:gluon}) is simple, the 
timelike functions are distinctly non--trivial.  Already from
Fig.~\ref{fig:quarkt} we observe a significant variation with 
the current quark mass. This indicates that the specific forms 
of timelike quark propagator functions possess pronounced
model and parameter dependences. Investigating the model dependence
requires changing the effective interaction~(\ref{eq:gluon})
and is not subject of the present study.  For the parameter 
dependence, we fortunately find (see below) that it does not 
significantly effect the model predictions for the meson masses.
There are divergences whenever $A$ and $B$ change sign at different $p^2$.  
Note that there are several particle--like poles ($p^2=-M^2(p^2)$) where the 
quark goes on--shell. This is an important observation, since it is 
not at all clear whether a confined quark can be on--shell. Indeed, 
one might be tempted to conclude that with on--shell singularities and 
a subsequent lack of confinement, the model will not be able to give 
meson masses at all. As we will see, this is fortunately not the case. 

\subsection{Mesons}

Turning to the mesons, we note that there are four model parameters, 
$\w,D,m_u$ and $m_s$ that we first have to fit to empirical data.  To 
this end, we initially choose the pseudoscalar meson observables 
$M_{\pi},M_{K}$ and $f_{\pi}$.  Then one parameter remains unconstrained 
by the pseudoscalar sector alone. However, the condition that the quark 
propagator function reflects dynamical chiral symmetry breaking, 
{\it i.e.} that $M(p^2=0)\approx 0.5{\rm GeV}$ leaves only a small window 
for the remaining choice, {\it cf.} table~\ref{tab:quark1}. 
All other masses and decay constants are subsequently model predictions.
The resulting model parameter and the predicted kaon decay constant $f_K$,
that is unexpectedly well reproduced, are shown in Table~\ref{tab:res1}. 
\begin{table}
\begin{center}
\begin{minipage}[t]{12.0cm}
\begin{tabular}{|c|c|c|c||c|c|c|c|}
$\w$ & $D$ & $m_u$ & $m_s$ & $M_{\pi}$ 
& $f_{\pi}$ & $M_K$ & $f_K$ \\ \hline
0.40 & 45.0 & $5\times10^{-3}$ & $0.120$ 
& 0.135 & 0.131 & 0.496 & 0.164 \\ \hline
0.45 & 25.0 & $5\times10^{-3}$ & $0.120$ 
& 0.135 & 0.131 & 0.496 & 0.163 \\ \hline
0.50 & 16.0 & $5\times10^{-3}$ & $0.115$ 
& 0.137 & 0.133 & 0.492 & 0.164 \\ \hline
\multicolumn{3}{|c}{experiment \cite{pdg}}& 
& 0.135 & 0.131 & 0.498 & 0.160
\end{tabular}
\end{minipage}
\end{center}
\caption{Parameter sets used and fit results for the
pseudoscalar mesons.  $M_{\pi}$, $f_{\pi}$ and $M_K$ are used
as input, $f_K$ is predicted. All units are ${\rm GeV}$.}
\label{tab:res1}
\end{table}
The subsequently predicted meson masses are shown in 
Tables~\ref{tab:res2}-\ref{tab:res5}. In all cases we have
the inequalities $M_{u\bar{u}}<M_{u\bar{s}}<M_{s\bar{s}}$,
where the subscript labels the flavor content. These relations
just reflect the quark--antiquark picture that is implicit
in the present Bethe--Salpeter approach.

Obviously both the pseudoscalar (table~\ref{tab:res1}) and vector 
mesons (table~\ref{tab:res2}) can be very well described within our 
model with the choice $\w\approx 0.5{\rm GeV}$. Our results agree with 
a previous analysis of the vector mesons based on an effective
interaction which included the perturbative type term~\cite{maris99}.
This shows that such terms do not have a large effect on the meson
masses, at least for the pseudoscalar and vector cases. Indeed, in
the context of low--energy meson phenomenology we conclude that 
the logarithmic tail, and its associated renormalization represent
an unnecessary obfuscation.
\begin{table}
\begin{center}
\begin{minipage}[t]{10.0cm}
\begin{tabular}{|c|c|c|c||c|c|c|}
$\w$ & $D$ & $m_u$ & $m_s$ & $M_{\rho}$ & $M_{K^*}$ & $M_{\phi}$ \\ \hline
0.40 & 45.0 & $5\times10^{-3}$ & $0.120$  & 0.748 & 0.939 & 1.072 \\ \hline
0.45 & 25.0 & $5\times10^{-3}$ & $0.120$ & 0.746 & 0.936 & 1.070 \\ \hline
0.50 & 16.0 & $5\times10^{-3}$ & $0.115$ & 0.758 & 0.946 & 1.078 \\ \hline
\multicolumn{3}{|c}{experiment \cite{pdg}}&
& 0.770 & 0.892 & 1.020
\end{tabular}
\end{minipage}
\end{center}
\caption{Results for the vector mesons. All units are ${\rm GeV}$.}
\label{tab:res2}
\end{table}

The situation for the scalar mesons (table~\ref{tab:res3}) is not quite
that clear. To begin with the particle data group~\cite{pdg} does not 
provide a clear picture in this channel but only quotes a wide range for 
the mass of the lowest scalar (0.4 -- 1.2 GeV). More detailed studies
of the pseudoscalar scattering amplitudes revealed that the assignment 
of the scalar meson nonet is not at all established~\cite{Syr1}. 
In particular, these mesons may not be simple quark--antiquark bound
states but {\it e.g.} might contain sizable admixture of 
2quark--2antiquark pairs~\cite{Ja77}. In that respect we might interpret
our results as a quark--antiquark model prediction for scalar mesons.
Our results suggest that such a picture is too simple for these mesons.
One might also speculate that the adopted ladder approximation could 
be insufficient.
\begin{table}
\begin{center}
\begin{minipage}[t]{10.0cm}
\begin{tabular}{|c|c|c|c||c|c|c|}
$\w$ & $D$ & $m_u$ & $m_s$ & $M_{u\ov{u}}$ 
& $M_{u\ov{s}}$ & $M_{s\ov{s}}$ \\ \hline
0.40 & 45.0 & $5\times10^{-3}$ & $0.120$  & 0.700 & 0.917 & 1.096 \\ \hline
0.45 & 25.0 & $5\times10^{-3}$ & $0.120$ & 0.675 & 0.908 & 1.099 \\ \hline
0.50 & 16.0 & $5\times10^{-3}$ & $0.115$ & 0.645 & 0.903 & 1.113
\end{tabular}
\end{minipage}
\end{center}
\caption{Results for the scalar mesons. The subscripts of $M$ 
denote the flavor content. All units are ${\rm GeV}$.}
\label{tab:res3}
\end{table}

For the axialvector mesons we have two channels that are 
distinguished by their charge conjugation properties, {\it cf.}
tables~\ref{tab:res4} and~\ref{tab:res5}. The quark--antiquark
pairs that are bound to axialvector modes with negative
charge conjugation eigenvalue tend to be lighter than those with 
the positive eigenvalue but otherwise equal quantum numbers. 
Generally we find that our predictions are lower than the 
assignments made by the particle data group~\cite{pdg}.
\begin{table}
\begin{center}
\begin{minipage}[t]{10.0cm}
\begin{tabular}{|c|c|c|c||c|c|c|}
$\w$ & $D$ & $m_u$ & $m_s$ & $M_{u\ov{u}}$ 
& $M_{u\ov{s}}$ & $M_{s\ov{s}}$ \\ \hline
0.40 & 45.0 & $5\times10^{-3}$ & $0.120$  & 0.804 & 0.994 & 1.128 \\ \hline
0.45 & 25.0 & $5\times10^{-3}$ & $0.120$ & 0.858 & 1.047 & 1.182 \\ \hline
0.50 & 16.0 & $5\times10^{-3}$ & $0.115$ & 0.912 & 1.098 & 1.230 \\ \hline
\multicolumn{3}{|c}{experiment \cite{pdg}}&                      
& 1.230 & 1.270 & 1.170 ?
\end{tabular}
\end{minipage}
\end{center}
\caption{Results for the axial-vector ($J^{PC}=1^{+-}$) mesons.  
The question mark indicates that the PDG did not assign the
charge conjugation property of the respective resonance.
All units are ${\rm GeV}$.}
\label{tab:res4}
\end{table}
\begin{table}
\begin{center}
\begin{minipage}[t]{10.0cm}
\begin{tabular}{|c|c|c|c||c|c|c|}
$\w$ & $D$ & $m_u$ & $m_s$ & $M_{u\ov{u}}$ 
& $M_{u\ov{s}}$ & $M_{s\ov{s}}$ \\ \hline
0.40 & 45.0 & $5\times10^{-3}$ & $0.120$  & 0.917 & 1.117 & 1.253 \\ \hline
0.45 & 25.0 & $5\times10^{-3}$ & $0.120$ & 0.918 & 1.124 & 1.270 \\ \hline
0.50 & 16.0 & $5\times10^{-3}$ & $0.115$ & 0.927 & 1.140 & 1.292 \\ \hline
\multicolumn{3}{|c}{experiment \cite{pdg}}&            
& 1.230 & 1.270 & 1.282
\end{tabular}
\end{minipage}
\end{center}
\caption{Results for the axial-vector ($J^{PC}=1^{++}$) mesons.  
All units are ${\rm GeV}$.}
\label{tab:res5}
\end{table}

We recognize from our results that the model predictions change 
only slightly within the large range of considered model parameters. 
This confirms that meson static properties are not too sensitive to the
conjectural parameter dependence of the timelike quark propagator 
functions. Presumably meson properties whose computation involves 
larger timelike will exhibit a stronger sensitivity.

Already from table~\ref{tab:res4} we observe that by increasing
$\omega$ the predicted mass of the $J^{PC}=1^{+-}$ meson with pion 
flavor quantum numbers approaches the empirical mass\footnote{In
all other channels the dependence on the parameters is much
softer within the allowed window.}. We therefore further increased
$\omega$ according to the rules discussed above. For $\w\sim0.8GeV$ we
reproduced the empirical value for the mass in that channel. However, this 
happened at the expense of significantly lowering $f_K$ and loosing
the proper description of the vector mesons. We recall that the parameter 
$\w$ has a physical interpretation as the location of the maximum of 
the interaction. Thus $\w=0.8GeV$ seems intuitively too large for 
low--energy hadron physics and an unsatisfactorily description of the 
$0^-$ and $1^-$ mesons comes without surprise.

For non--diagonal flavor structures such as $u\bar{s}$, charge 
conjugation actually is not a sensible quantum number and the 
corresponding axial vector mesons $1^{++}$ and $1^{+-}$ may mix.
In table~\ref{tab:res6} we present the results obtained from the full 
calculation that combines the Dirac decompositions~(\ref{eq:axip}) 
and~(\ref{eq:axim}).
\begin{table}
\begin{center}
\begin{minipage}[t]{10.0cm}
\begin{tabular}{|c|c|c|c||c|c|c|}
$\w$ & $D$ & $m_u$ & $m_s$ & $M_{u\ov{u}}$
& $M_{u\ov{s}}$ & $M_{s\ov{s}}$ \\ \hline
0.40 & 45.0 & $5\times10^{-3}$ & $0.120$ & 0.807 & 0.990 & 1.131 \\ \hline
0.45 & 25.0 & $5\times10^{-3}$ & $0.120$ & 0.861 & 1.040 & 1.185 \\ \hline
0.50 & 16.0 & $5\times10^{-3}$ & $0.115$ & 0.915 & 1.085 & 1.233 \\ \hline
\end{tabular}
\end{minipage}
\end{center}
\caption{Results for the axial-vector mesons allowing for mixing 
of the Dirac structures in eqs.~(\protect\ref{eq:axip}) 
and~(\protect\ref{eq:axim}). All units are ${\rm GeV}$.}
\label{tab:res6}
\end{table}
Since our Bethe--Salpeter formalism only yields the lowest 
mass eigenstate within a given channel, the results presented
in table~\ref{tab:res6} should be compared to those in 
table~\ref{tab:res4}. The tiny changes for the flavor 
diagonal mesons are numerical artifacts. Surprisingly the
changes for the non--diagonal flavor structure are also
only of the order 1\%. This suggests only a small mixing
between the $1^{++}$ and $1^{+-}$ states with the flavor
structure $u\bar{s}$ and the $1^{++}$ and $1^{+-}$ channels 
represent good approximations to the actual eigenstates.

\begin{figure}
\centerline{
\epsfig{figure=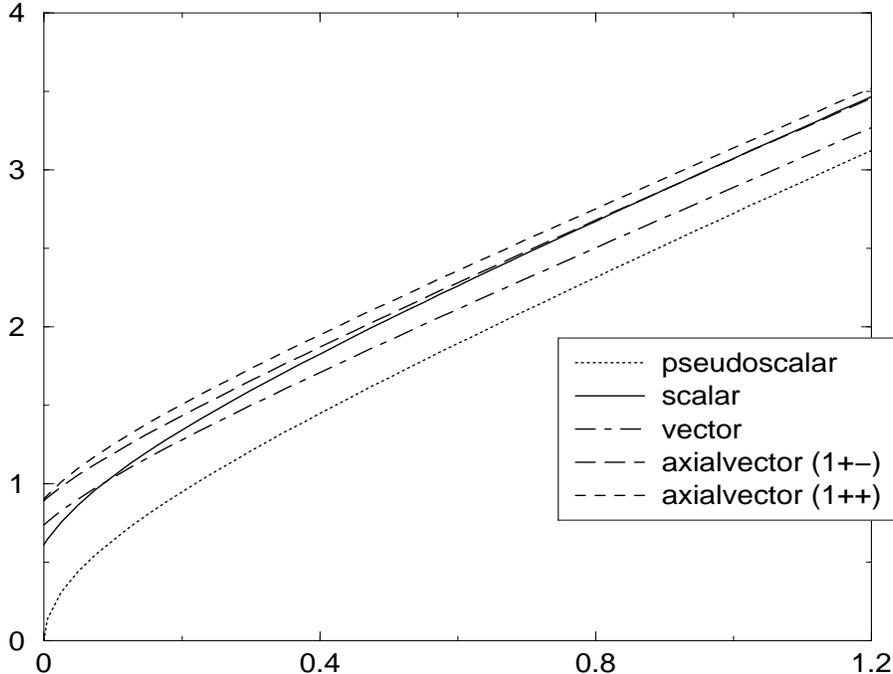,width=12cm,height=9cm}}
\caption{Meson masses as a function of the (equal) quark mass, 
$m_0$.  $\w=0.5GeV, D=16GeV^{-2}$.  All units are ${\rm GeV}$.}
\label{fig:mass1}
\end{figure}

We can extend our model beyond the light flavors up, down 
and strange. The only modification is the increase of the current quark 
mass, $m_0$. In Fig.\ \ref{fig:mass1} we show how the meson 
masses increase as the (equal) quark masses are increased into the 
charm sector $m_c=1.125{\rm GeV}$.  
Exactly the same numerical code is used to construct these solutions to 
the Schwinger--Dyson and Bethe--Salpeter equations as for the light 
flavors. Clearly seen is the smooth way the masses increase from the 
chiral limit ($m_0=0$) into the heavy quark sector ($m_0=m_c$). This 
represents a convincing indicator for the stability of our technique. The 
$c\ov{c}$-meson masses can be loosely extracted (table~\ref{tab:res7}) 
and the data are surprisingly well reproduced. 
\begin{table}[t]
\begin{center}
\begin{minipage}[t]{6.5cm}
\begin{tabular}{|c|c|c|}
$J^{P(C)}$ & $M_{c\ov{c}}$ & experiment \cite{pdg}\\ \hline
$0^-$ & 2.97 & $\eta_c$: 2.98 \\ \hline
$1^-$ & 3.13 & $J/\psi$: 3.10 \\ \hline
$0^+$ & 3.32 & $\chi_{c0}$: 3.42\\ \hline
$1^{++}$ & 3.38 & $\chi_{c1}$: 3.51 \\ \hline
$1^{+-}$ & 3.31 & ?
\end{tabular}
\end{minipage}
\end{center}
\caption{Results for the $c\ov{c}$-meson states.  
$\w=0.4,D=45.0,m_c=1.125$.  $m_c$ is fitted approximately from the 
$\eta_c$ mass.}
\label{tab:res7}
\end{table}
We did not expect to be able to describe such a heavy 
system with such a simple model, derived (and fitted) as it is from 
pion physics.  The lack of the correct UV behavior for the gluon is 
seemingly at odds with the scales present.  However, the present results 
suggest that the Bethe--Salpeter equation is capable of describing 
{\it all} the angular momentum states equally well in the charm quark sector.

\section{Summary and Outlook}

In this paper we have studied the low--lying mesons as quark--antiquark 
bound states in a covariant approach using an effective interaction.
This interaction is characterized by gluon exchange with the gluon 
propagator being dressed by a Gaussian shape function. The interaction 
is completed by the quark--gluon vertex that we take to be the tree--level 
perturbative 
one. In this manner the rainbow--ladder approximation to the system
of Schwinger--Dyson and Bethe--Salpeter equation accounts for chiral 
symmetry. With this effective interaction, we have then consistently 
treated this system of integral equations by precisely implementing 
the quark propagator functions that solve the Schwinger--Dyson equations 
into the Bethe--Salpeter equations. Once the effective interaction
exceeds a certain strength, the Schwinger--Dyson equations exhibit 
dynamical chiral symmetry breaking and the pseudoscalar mesons emerge 
as {\it would--be} Goldstone bosons. We have then used observed properties 
of the pseudoscalar mesons to determine the model parameters. The kaon 
decay constant represents a model prediction. It turned out to be in 
good agreement with the empirical data. Furthermore our results for the 
vector meson masses match the experimental data. The situation in the 
scalar channel is less satisfying. As we solely consider the mesons 
as bound states of quark--antiquark pairs, it is not surprising that 
the mass eigenvalues increase with the strangeness content. On the other 
hand it is astonishing that for current quark masses, $m_0\ge0.2{\rm GeV}$, 
the lightest scalar mesons turn out to be heavier than the lightest 
vector mesons. When discussing these results it must be noted that the role 
of the scalar mesons is still under intense debate. In particular, the 
question whether they should indeed be considered as quark--antiquark 
bound states is not yet completely resolved. There are 
indications, see e.g.\ Ref.\ \cite{Syr1} and references therein,
that the scalar meson masses should 
actually decrease with the strangeness content of these mesons. This can
be understood if these mesons are considered as 2quark--2antiquark 
bound states in the sense of diquark--antidiquark systems~\cite{Ja77}. 

As an outlook we mention that there is an elegant way to extend the 
present model to incorporate such degrees of freedom. The Bethe--Salpeter 
treatment can be straightforwardly extended to study bound states of 
diquark--antidiquark pairs, once a binding mechanism is established. This 
could either be achieved by a gluon exchange similar to eq.~(\ref{eq:gluon})
or by quark exchange between a quark and a diquark. The latter
approach has been intensively studied and the corresponding
vertex is known from modeling baryon properties~\cite{tuediq}.
It will also be interesting to see whether these additional degrees of 
freedom will also affect the mass predictions for the axialvector mesons
that currently tend to be on the low side. Investigations in this 
direction are in progress.

Finally we would like to repeat that the stability of our
treatment -- measured by the actual momentum partition invariance --
allows us to even describe mesons with charm quantum numbers.
Without further modifications of the model parameter, except
the corresponding current quark mass, our approach reproduces
the mass eigenvalues of the flavor neutral states unexpectedly
well.

\section*{Acknowledgments} 

\noindent
We thank J.C.R.~Bloch, C.~S.~Fischer, D.~J.~L\"ange, P.~Maris, M.~Oettel, 
H.~Reinhardt, S.~M.~Schmidt, and P.~C.~Tandy for valuable discussions.
\noindent
This work has been supported by 
DFG~(Al-297/3--3, Al-297/3--4, We-1254/3--2, We-1254/4--2)
and  COSY~(contract no. 41376610).


\goodbreak

\appendix
\section{Relativistic covariance}

\begin{figure}
\centerline{
\epsfig{figure=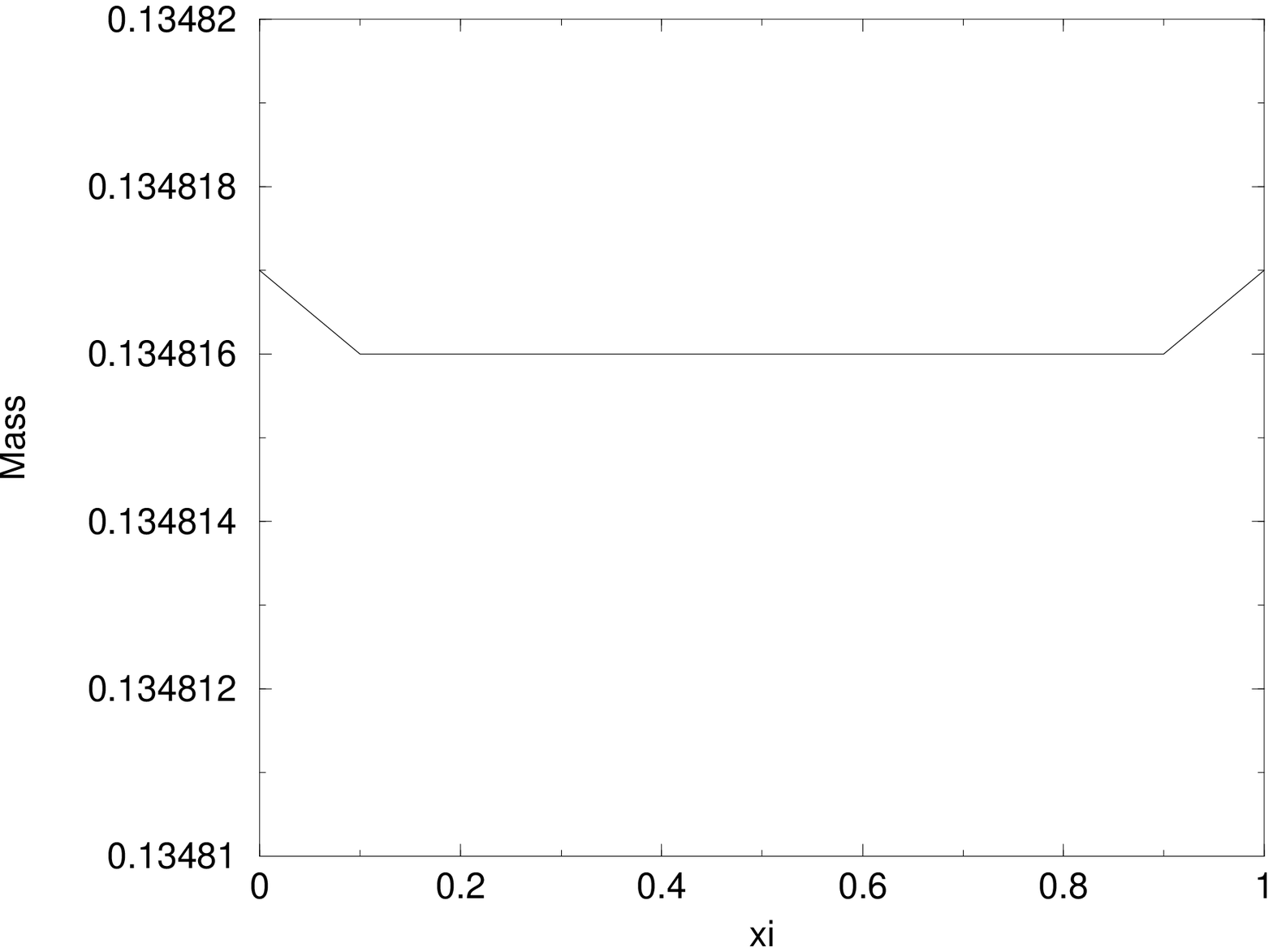,width=6.5cm,height=6cm}\hspace{2cm}
\epsfig{figure=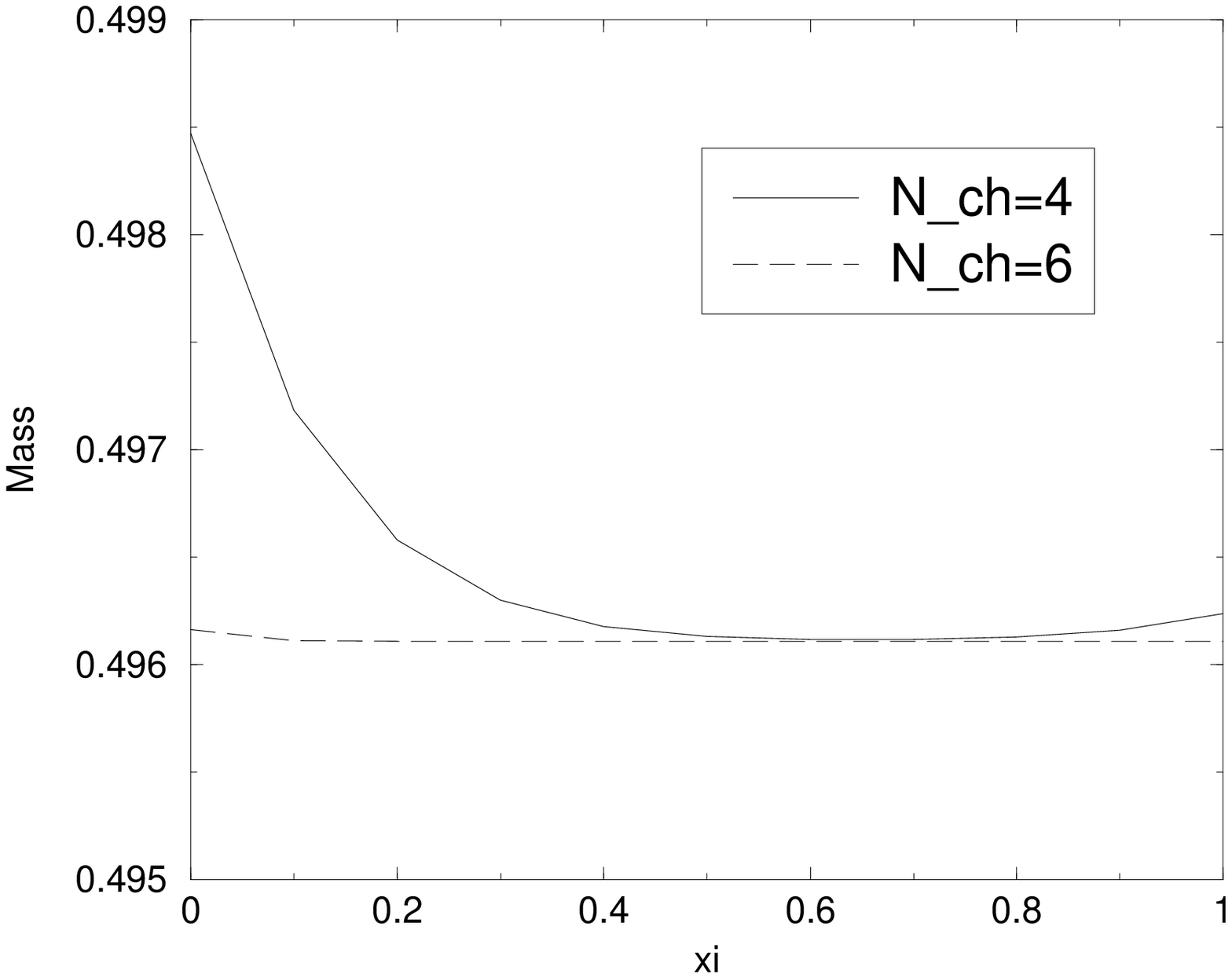,width=6.5cm,height=6cm}}
\caption{Dependence of the pion (left) and kaon (right) masses on
the momentum partition $\xi$.  Model parameters are:
$\w=0.4$, $D=45$, $m_u=5\times10^{-3}$ and $m_{s}=0.120$.
All units are~${\rm GeV}$. For the calculation of the pion mass
$N_{\rm ch}=4$ has been used.}
\label{fig:xi1}
\end{figure}

\begin{figure}
\centerline{
\epsfig{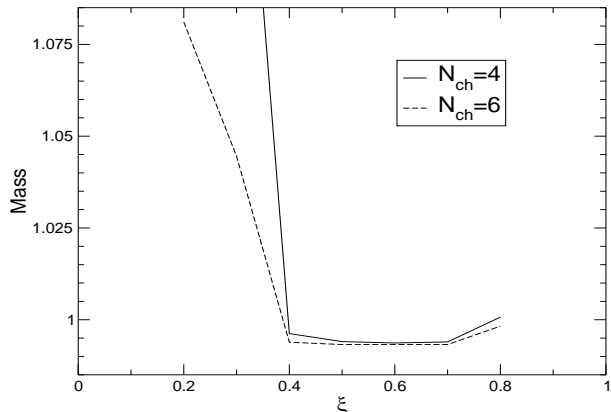}}
\caption{Dependence of the axial-vector ($J^{PC}=1^{+-}$) mass on the
momentum partition $\xi$.  One set of parameters has been used:
$\w=0.4$, $D=45$, $m_u=5\times10^{-3}$ and $m_{s}=0.120$.
All units are~${\rm GeV}$.}
\label{fig:xi2}
\end{figure}

As mentioned in section III the truncation of taking $N_{\rm ch}$ finite 
in eq.~(\ref{eq:cheb}) violates relativistic covariance. Studying 
the dependence of our results on  the momentum partitioning parameter $\xi$
we can provide a measure of these violations originating from the numerical
method. The results for the 
pseudoscalar case are shown in Fig.\ \ref{fig:xi1}.  
The pion mass 
is extremely stable over the whole range $\xi\in[0,1]$, whereas for 
the kaon, the asymmetry between the quark masses must be compensated 
for by increasing the number of Chebyshev moments, $N_{\rm ch}$.  
Clearly, for the light pseudoscalars, the technique of inserting the quark 
propagators directly results in extremely stable results. In the case 
of the other mesons, the stability becomes less evident as one is 
studying heavier mass resonances (but with the same quarks).  As an 
example, we display the axial--vector ($J^{PC}=1^{+-}$) $u\ov{s}$ meson 
(Fig.\ \ref{fig:xi2}).  
There is still a clear range of stability 
$\xi\in[0.4,0.7]$.  Outside this range however, we notice that the 
technique does not simply break down (a mass can still be found) -- 
it merely gets numerically more difficult to get trustworthy results.
We assert the increasing dependence on $\xi$ with larger
meson masses to the fact that with larger meson masses the 
calculation becomes more sensitive to the violent behavior seen in
the timelike quark propagator functions. The larger the influence 
of that behavior to the calculation the higher we have to assume
the number of Chebyshev moments for an adequate representation.
Away from this timelike regime the quark propagator functions
are smooth. This results in a good representation of the 
Bethe--Salpeter kernel and vertex functions with only a few
Chebyshev moments.


\end{document}